\documentclass[epj]{webofc}
\usepackage[utf8]{inputenc}
\usepackage[varg]{txfonts}   
\usepackage{booktabs}
\usepackage{xcolor}
\definecolor{darkred}{rgb}{0.4,0.0,0.0}
\definecolor{darkgreen}{rgb}{0.0,0.4,0.0}
\definecolor{darkblue}{rgb}{0.0,0.0,0.4}
\usepackage[bookmarks,linktocpage,colorlinks,
    linkcolor = darkred,
    urlcolor  = darkblue,
    citecolor = darkgreen]{hyperref}
\usepackage{ytableau}
\usepackage{bm}
\usepackage{braket}
\wocname{EPJ Web of Conferences}
\woctitle{Lattice2017}
\newcommand{\tr}{\operatorname{tr}}
\ytableausetup{centertableaux}
\begin{document}
%
\selectlanguage{english}
\title{%
	Double-winding Wilson loops in $SU(N)$ Yang-Mills theory
}
\subtitle{-- A criterion for testing the confinement models --}
\author{%
\firstname{Ryutaro} \lastname{Matsudo}\inst{1}\fnsep\thanks{Speaker, \email{afca3071@chiba-u.jp}.} \and
\firstname{Kei-Ichi} \lastname{Kondo}\inst{2} \and
\firstname{Akihiro}  \lastname{Shibata}\inst{3}
}
\institute{%
Department of Physics, Faculty of Science and Engineering, Chiba University, Chiba 263-8522, Japan
\and
Department of Physics, Faculty of Science, Chiba University, Chiba 263-8522, Japan
\and
Computing Research Center, High Energy Accelerator Research Organization (KEK), Tsukuba 305-0801, Japan
}
\abstract{%
We examine how the average of double-winding Wilson loops depends on the number of color $N$ in the $SU(N)$ Yang-Mills theory. 
In the case where the two loops $C_1$ and $C_2$ are identical, we derive the exact operator relation which relates the double-winding Wilson loop operator in the fundamental representation to that in the higher dimensional representations depending on $N$.
By taking the average of the relation, we find that the difference-of-areas law for the area law falloff recently claimed for $N=2$ is excluded for $N \geq 3$, provided that the string tension obeys the Casimir scaling for the higher representations. 
In the case where the two loops are distinct, we argue that the area law follows a novel law $(N - 3)A_1/(N-1)+A_2$ with $A_1$ and $A_2 (A_1<A_2)$ being the minimal areas spanned respectively by the loops $C_1$ and $C_2$, which is neither sum-of-areas ($A_1+A_2$) nor difference-of-areas ($A_2 - A_1$) law when ($N\geq3$). Indeed, this behavior can be confirmed in the two-dimensional $SU(N)$ Yang-Mills theory exactly.
}
\maketitle

\section{Introduction}\label{intro}
\begin{figure}[t]
\centering
\includegraphics[width=0.2\hsize]{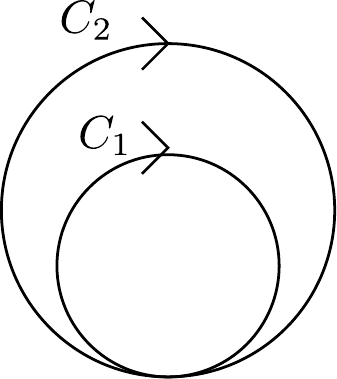}
\caption{A double-winding Wilson loop.}
\label{contour_double}
\end{figure}

Which degrees of freedom are responsible for confinement?
This question is not yet satisfactorily answered but there are two promising candidates, center vortices and \textit{Abelian} monopoles.
Recently, a criterion for testing these two candidates in the $SU(2)$ Yang-Mills theory has been proposed in \cite{Greensite:2014gra}.
It is a double-winding Wilson loop, which is
defined as a Wilson loop consisting of two coplanar loops $C_1$ and $C_2$, where $C_1$ lies entirely in the minimal area of $C_2$ and the two loops share one point and have the same direction,
as indicated in Fig.\ \ref{contour_double}.
The average of a double-winding Wilson loop is expected to depend on the model.
In the center vortex model, the average of a double-winding Wilson loop decreases exponentially with the difference of areas.
This behavior is called the difference-of-areas law.
In the models associated with the Abelian monopoles, it decreases exponentially with the sum of areas.
This behavior is called the sum-of-areas law.
Thus we can test the two models by investigating the true behavior of the average of the double-winding Wilson loop.
Indeed, the difference-of-areas behavior is supported by the lattice simulations.

The main purpose of this talk is to extend this argument for the $SU(2)$ gauge group to the $SU(N)$ gauge group with an arbitrary $N\geq 3$.
As the first step, we try to find the true behavior of double-winding Wilson loops.
Actually, in the $SU(N)$ case, if $C_1$ and $C_2$ are identical, we can determine the behavior of the averages of double-winding Wilson loops by using group identities and assuming the Casimir Scaling.
This fact is already mentioned in \cite{Greensite:2014gra} in the $SU(2)$ case.
In addition, under a reasonable assumption, we can determine the behavior even if $C_1$ and $C_2$ are not identical.
Consequently, we find that in the $SU(N)$ ($N\geq3$) case, the average of a double-winding Wilson loop does not obey the law similar to that in the $SU(2)$ case, and therefore investigating double-winding Wilson loops is not necessarily appropriate to test the models.

This fact leads us to investigate general $m$-times-winding Wilson loops by following the similar procedures.
As a result, we see that we should use $N$-times-winding Wilson loops to test the models.

In the next section, first, we explain how we analyse the averages of double-winding Wilson loops when $C_1$ and $C_2$ are identical.
Next, we explain how we extend our analysis to the case where $C_1$ and $C_2$ are not identical.
Lastly, we explain why it is not appropriate for the $SU(N)$ group.
In Section \ref{mTimes}, we discuss the general $m$-times-winding case, and conclude that the $N$-times-winding Wilson loop is appropriate for the $SU(N)$ Yang-Mills theory. 

\section{Double-winding Wilson loops} \label{double}
In this section, we discuss the double-winding Wilson loop.
First, we investigate the average of the double-winding Wilson loop when $C_1$ and $C_2$ are identical by using the group identity and assuming the Casimir scaling.
Next, we discuss the case where $C_1$ and $C_2$ are not identical by assuming the factorization of coplanar non-overlapping loops.

In the $SU(2)$ case, the expectation value of a double-winding Wilson loop with identical subloops can be estimated by using the group identity
\begin{align}
	\tr U^2 = \tr U_A - 1,
\end{align}
where $U$ is an arbitrary element of the group in the fundamental representation and $U_A$ is the image of $U$ under the adjoint representation.
Representing $U$ in terms of the phase factor, taking the average, and assuming exponential falloff of the Wilson loop in the adjoint representation, we can see that the average approaches a negative constant as the size increases.

Similarly, in the $SU(N)$ case, we can also estimate it by using the group identity
\ytableausetup{boxsize=0.5em}
\begin{align}
	\tr U^2 = \tr U_{\ydiagram{2}} - \tr U_{\ydiagram{1,1}},
\end{align}
where the Young diagrams denote the representations,
\ytableausetup{boxsize=0.7em}
\begin{align}
	\ydiagram{2}&= [2,0,\ldots,0] = \bm{N(N+1)/2} \\
	\ydiagram{1,1}&=[0,1,0,\ldots,0] = \bm{N(N-1)/2}.
\end{align}
The proof of this formula is given in \cite{Matsudo:2017wyv}.
Using this group identity the double-winding Wilson loop with identical loops can be written as
\ytableausetup{boxsize=0.5em}
\begin{align}
	W_2 = \frac{N+1}2 W_{\ydiagram{2}} - \frac{N-1}2 W_{\ydiagram{1,1}}.
\end{align}

By using this relation and assuming the Casimir scaling, we can estimate the expectation value for the loop of intermediate size.
The ratios of the values of the quadratic Casimir in the above representations to that in the fundamental representation are given by
\ytableausetup{boxsize=0.7em}
\begin{align}
	&C_2\left(\ydiagram{2}\right)/C_2(F) = \frac{2(N+2)}{N+1},\\
	&C_2\left(\ydiagram{1,1}\right)/C_2(F) = \frac{2(N-2)}{N-1}.
\end{align}
By assuming the Casimir scaling, therefore, in the case that the minimal area $S$ is intermediate, we obtain
\begin{align}
	\braket{W_2} \simeq a_N\exp\left(-2\frac{N+2}{N+1}\sigma_FS\right) -b_N\exp\left( -2\frac{N-2}{N-1}\sigma_F S \right),\quad a_N,b_N>0, \label{2WindResult}
\end{align}
where $a_N$ and $b_N$ are positive constants depending on $N$.
For sufficiently large $S$ the dominant part is the second term
because $(N+2)/(N+1)$ is larger than $(N-2)/(N-1)$.
For $N\geq3$, this behavior is consistent with neither difference-of-areas law nor sum-of-areas law.

\begin{figure}[t]
\centering
\includegraphics[width=0.5\hsize]{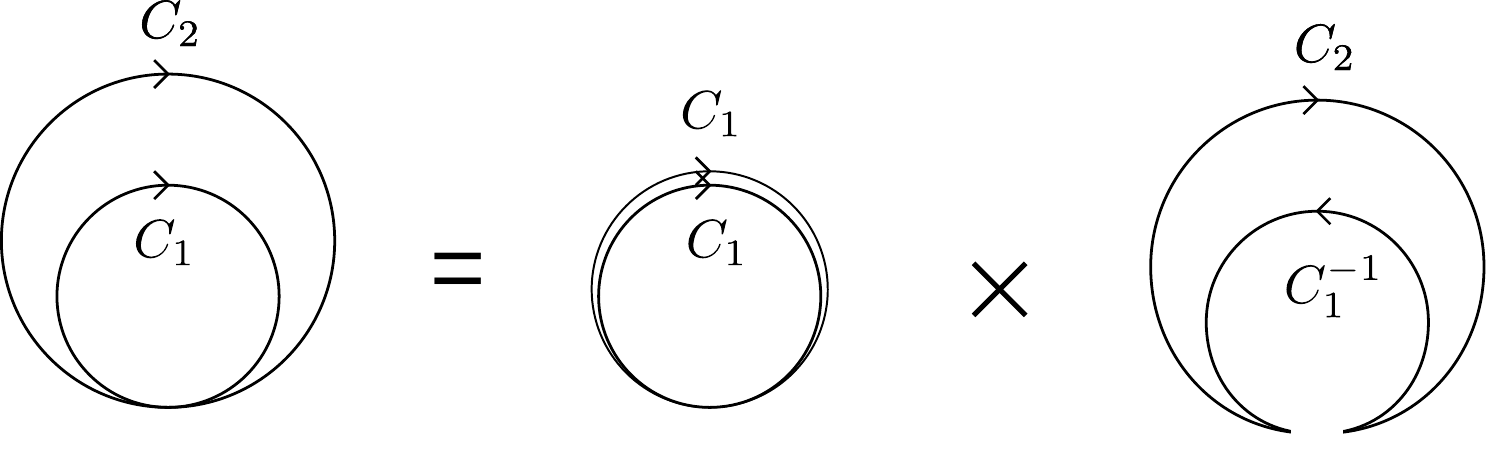}
\caption{Decomposition of the loop we used.}
\label{contour_double_identical+difference}
\end{figure}

Even if $C_1$ and $C_2$ are not identical, we can determine the behavior by assuming the factorization of coplanar non-overlapping loops.
Here the term ``non-overlapping'' means that the intersection of the minimal areas of two loops is empty.
This assumption is not so strange because this is consistent with the usual area law.
By using this assumption, we can decompose a double-winding Wilson loop into the double-winding Wilson loop with the loop $C_1^2$ and the single-winding Wilson loop with the loop $C_1^{-1}C_2$.
\begin{align}
	\braket{\tr U(C_1\times C_2)} = \braket{\tr U(C_1^2) U(C_1^{-1}\times C_2)} \simeq \tr\braket{U(C_1^2)}\braket{U(C_1^{-1}\times C_2)},
\end{align}
as indicated in Fig.\ \ref{contour_double_identical+difference}.
By substituting the result for the identical subloops and usual single winding case Eq.\ (\ref{2WindResult}), we obtain 
\begin{align}
	\braket{W(C_1\times C_2)} &\simeq 
	c\exp\left(-\sigma_F\left( \frac{N+3}{N+1}A_1 + A_2 \right)\right) 
	-d\exp\left(-\sigma_F\left( \frac{N-3}{N-1}A_1 + A_2 \right)\right), \label{2windResult}
\end{align}
where $c$ and $d$ are positive constants and $A_1$ and $A_2$ are the minimal areas of $C_1$ and $C_2$ respectively.
This result is consistent with the case of $A_1=A_2$ and that of $A_1=0$.

In two-dimensional spacetime, we can calculate it exactly as
\begin{align}
	\braket{W_2} = &\frac{N+1}2 \exp\left(-\frac{g^2}2\frac{N^2-1}{2N}\left(\frac{N+3}{N+1}A_1+A_2 \right)\right) 
	- \frac{N-1}2 \exp\left(-\frac{g^2}2\frac{N^2-1}{2N}\left(\frac{N-3}{N-1}A_1+A_2\right)\right),
\end{align}
following \cite{Bralic:1980ra}.
Therefore the above result is confirmed at least in two-dimensional spacetime.

Now let us discuss whether the behavior Eq.\ (\ref{2windResult}) is appropriate to test the models.
If $A_1$ is sufficiently large, the dominant part is the second term.
For example, in $N=3$ case, the dominant part is
\begin{align}
	\braket{W(C_1\times C_2)} \simeq -d \exp\left(-\sigma_F A_2\right),
\end{align}
which means that the average does not decrease exponentially with $A_1$.
However in the models associated with Abelian monopoles, the average is expected to decrease exponentially with $A_1$ according to \cite{Greensite:2014gra}.
In this case, therefore, we can use double-winding Wilson loop to discriminate the models.
However in $N\geq 4$ case, the double-winding Wilson loop is not appropriate to our purpose.
This is because, according to Eq.\ (\ref{2windResult}), the average decreases exponentially with both $A_1$ and $A_2$, and also in the models associated with Abelian monopoles, the similar behavior is expected.
This means that to distinguish these two behavior, we need to know the detailed information about the ``string tension''.

\section{General $m$-times-winding Wilson loops} \label{mTimes}
In the previous section, we have shown that for arbitrary $SU(N)$ gauge group, double-winding Wilson loops are not necessarily appropriate to test the models.
We can more easily clarify the difference of the models by using the Wilson loop which winds $N-1$ times around $C_1$ and once around $C_2$.
To see this, we consider the general $m$-times-winding Wilson loop with identical subloops.
Also in this case, we can estimate its average by using the group identity.

The trace of the $m$th power of a group element $U$ can be rewritten using the trace of the group elements in the higher dimensional representations:
for $m<N$
\begin{align}
	\tr U^m =
	U_{\scalebox{0.5}{
		\ytableaushort{{} {} {\none[\cdot]}{\none[\cdot]}{\none[\cdot]}{}}
	}}
	-
	U_{\scalebox{0.5}{
		\ytableaushort{{} {} {\none[\cdot]}{\none[\cdot]}{\none[\cdot]}{},
		{}}
	}}
	+ \cdots + (-1)^{\ell-1}
	U_{\scalebox{0.5}{
		\ytableaushort{{} {} {\none[\cdot]}{\none[\cdot]}{\none[\cdot]}{},
		{},
		{\none[\cdot]},
		{\none[\cdot]},
		{\none[\cdot]},
		{}}
	}}
	+ \cdots + (-1)^{m-1}
	U_{\scalebox{0.5}{
	\ytableaushort{{},{},{\none[\cdot]},{\none[\cdot]},{\none[\cdot]},{}}
	}},
\end{align}
where all diagrams are "L-shape" or "I-shape", there are $m$ boxes in each diagram, and there are $\ell$ raws in the diagram in $\ell$th term,
and for $m\geq N$
\begin{align}
	\tr U^m = 
	U_{\scalebox{0.5}{
		\ytableaushort{{} {} {\none[\cdot]}{\none[\cdot]}{\none[\cdot]}{}}
	}}
	-
	U_{\scalebox{0.5}{
		\ytableaushort{{} {} {\none[\cdot]}{\none[\cdot]}{\none[\cdot]}{},
		{}}
	}}
	+ \cdots + (-1)^{\ell-1}
	U_{\scalebox{0.5}{
		\ytableaushort{{} {} {\none[\cdot]}{\none[\cdot]}{\none[\cdot]}{},
		{},
		{\none[\cdot]},
		{\none[\cdot]},
		{\none[\cdot]},
		{}}
	}}
	+ \cdots + (-1)^{N-1}
	U_{\scalebox{0.5}{
		\ytableaushort{{} {} {\none[\cdot]}{\none[\cdot]}{\none[\cdot]}{},
		{},
		{\none[\cdot]},
		{\none[\cdot]},
		{\none[\cdot]},
		{}}
	}},
\end{align}
where there are $N$ terms.
The proof of this formula is given in \cite{Matsudo:2017wyv}.

Using this formula and assuming the Casimir scaling, for the loop of an intermediate size, we can determine the dominant term for the average of the $m$-times-winding Wilson loop with identical subloops as
\begin{align}
	\braket{W_m} \simeq \begin{cases}
		(-1)^{m-1}a_{Nm}\exp\left( -\frac{m(N-m)}{N-1}\sigma_F S\right)  &\text{for }m<N, \\
		\mathcal (-1)^{N-1}a_{NN}&\text{for }m=N, \\
		(-1)^{m-1}a_{Nm}\exp\left( -\frac{m(m-N)}{N+1}\sigma_F S\right) &\text{for }m>N.
	\end{cases}
\end{align}
The reason why $\braket{W_N}$, the expectation value $\braket{W_m}$ for $m=N$, does not decrease exponentially is as follows.
The Young diagram
\begin{align}
	\ytableaushort{{},{},{\none[\cdot]},{\none[\cdot]},{\none[\cdot]},{}}
\end{align}
corresponds to the trivial representation if the total number of the boxes is $N$.
Hence, if $m=N$, the last term is equal to $(-1)^{N-1}$.
In this way we see that $\braket{W_N}$ does not follow the area law.
It seems that we can use $W_N$ in the $SU(N)$ case in the similar way to the double-winding Wilson loop in $SU(2)$ case.

By using the result, we can construct Wilson loops whose averages exhibit the difference-of-areas behavior.
One of these is the Wilson loop which winds $N-1$ times around $C_1$ and once around $C_2$.
In the same way as the double-winding case, assuming the factorization of coplanar non-overlapping loops, we obtain
\begin{align}
	\braket{W(C_1^{N-1} \times C_2)} \simeq (-1)^{N-1}c \exp\left( -\sigma_F(A_2-A_1)\right), \label{newWil}
\end{align}
where $c$ is a positive constant.

To see the Wilson loop Eq.\ (\ref{newWil}) can be used to test the models, let us simply discuss the expected behavior of the average in the center vortex model and the models associated with Abelian monopoles.
First in the center vortex model, if the loop is sufficiently large so that the thickness of center vortices is ignored,
the difference-of-areas law is expected to hold.
The reason is as follows.
If a center vortex pierces $A_1$, it links the loop $N$ times and hence the phase factor is multiplied by the $N$th power of an element of center, which is equal to one.
This means there are no effects in this case.
If a center vortex pierces $A_2-A_1$, it links the loop once and hence the effect is the same as the usual single-winding case.
Therefore the average is same as that of the single-winding Wilson loop with the minimal area $A_2-A_1$.
Next we consider the models associated with Abelian monopoles.
Let the loop be rectangular.
Then the Wilson loop Eq.\ (\ref{newWil}) represents the situation that $N$ quark-antiquark pairs with infinitely large masses are created at a time and annihilated a long time later.
In the dual superconductivity picture, the electric flux forms $N$ tubes in this situation and the total energy is the sum of the energy of each flux tube.
This means that the average of the Wilson loop decreases exponentially with both $A_1$ and $A_2$.
This is explicitly distinguishable from the true behavior.
Notice that this argument is valid only if we neglect the effect of the off-diagonal gluons.
This is because the off-diagonal gluons can neutralize the pair of the quark and the antiquark even if the off-diagonal gluons have large mass, and therefore the strings connecting the quark and the antiquark can break.

Next let us consider the case that the loops has asymptotically large size.
In this case, we should assume the $N$-ality dependence of the string tension instead of the Casimir scaling.
Then we can estimate the average of the $m$-times-winding Wilson loop with identical subloops as
\begin{align}
	\braket{W_m} \simeq
	(-1)^{m-1} a_{Nm}\exp (-\sigma_k S), \quad m\equiv k\, (\mathrm{mod}\,N) \quad k\leq N,
\end{align}
where $\sigma_k$ is the asymptotic string tension of a single-winding Wilson loop in the representation whose $N$-ality is $k$.
For $m> N$, the loop of intermediate size behaves differently from the loop of asymptotic size.
Especially, $m=nN$ ($n\in\mathbb N$) case is rather special because for the loop of asymptotic size the average of $nN$-times-winding Wilson loops with identical subloops does not exhibit the area law, while the loops of intermediate size obeys the area law.
In this case, the above argument in the center vortex model provides the result for the loop of asymptotic size, which is different from that for the loop of intermediate size.
This means that in this case we need to include the effect of thickness of center vortices, i.e.\ the effect of the off-diagonal gluons.
This fact suggests that we cannot neglect the off-diagonal components of the gauge field to consider the quark confinement problem in either cases.

%
%
%
%
%

\section*{}
\textbf{Acknowledgements}: R. M. was supported by Grant-in-Aid for JSPS Research Fellow Grant Number 17J04780. K.-I. K. was supported by Grant-in-Aid for Scientific Research, JSPS KAKENHI Grant Number (C) No.15K05042.

\bibliography{lattice2017}

\end{document}